# A Performance Study of GA and LSH in Multiprocessor Job Scheduling

**Mrs S. R. Vijayalakshmi [1] and Dr G. Padmavathi [2]**

**[1]. Assistant Professor, School of Information Technology and Science, Dr.G.R.D College of Science, Coimbatore, Tamil Nadu, India**

**[2].Professor and Head, Dept.of Computer Science, Avinashilingam University for Women, Coimbatore – 43, Tamil Nadu, India.**

## Abstract

Multiprocessor task scheduling is an important and computationally difficult problem. This paper proposes a comparison study of genetic algorithm and list scheduling algorithm. Both algorithms are naturally parallelizable but have heavy data dependencies. Based on experimental results, this paper presents a detailed analysis of the scalability, advantages and disadvantages of each algorithm. Multiprocessors have emerged as a powerful computing means for running real-time applications, especially where a uni-processor system would not be sufficient enough to execute all the tasks. The high performance and reliability of multiprocessors have made them a powerful computing resource. Such computing environment requires an efficient algorithm to determine when and on which processor a given task should execute. In multiprocessor systems, an efficient scheduling of a parallel program onto the processors that minimizes the entire execution time is vital for achieving a high performance. This scheduling problem is known to be NP-Hard. In multiprocessor scheduling problem, a given program is to be scheduled in a given multiprocessor system such that the program's execution time is minimized. The last job must be completed as early as possible. Genetic algorithm (GA) is one of the widely used techniques for constrained optimization. Genetic algorithms are basically search algorithms based on the mechanics of natural selection and natural genetics. List scheduling techniques assign a priority to each task to be scheduled then sort the list of tasks in decreasing priority. As processors become available, the highest priority task in the task list is assigned to be processed and removed from the list. If more than one task has the same priority, selection from among the candidate tasks is typically random. This paper compares Genetic algorithm (GA) with List Scheduling heuristic (LSH) to solve scheduling problem of multiprocessors.

**Keywords:** *Task scheduling · Parallel computing · Heuristic algorithms. Real time systems.*

## 1. Introduction

Scheduling a set of dependent or independent tasks for parallel execution on a set of processors is an important and computationally complex problem. Parallel program can be decomposed into a set of smaller tasks that generally have dependencies. The goal of task scheduling is to assign tasks to available processors such that precedence requirements between tasks are satisfied and the overall time required to execute all tasks, the *makespan*, is minimized. Various studies have proven that finding an optimal schedule is an NP-complete problem even in the simplest forms. As finding an optimal solution is not feasible, a large number of algorithms were proposed which attempt to obtain a near-optimal solution for various variants of the multiprocessor task scheduling problem. These algorithms usually trade the computational complexity of the scheduling algorithm itself to the quality of the solution. Algorithms based on complex, iterative search can usually (but not always) outperform simple one-pass heuristics, but their computational complexity makes them less scalable. The comparison of the various approaches is made difficult by the lack of an agreed benchmark problem, and the variety of assumptions made by the developers. This paper compares the performance of genetic algorithms with list scheduling heuristics. The rest of the paper is organized as follows. The next section presents related research work. The structure of the problem and assumptions are described in Section 3. The considered scheduling algorithms discussed in detail in Section 4. Section 5 presents simulation results and analysis. Conclusions are offered in Section 6.

## 2. Related Work

In order to allocate parallel applications to maximize throughput, task precedence graph (TPG) or a task interaction graph (TIG) are modeled.

The system usually schedules tasks according to their deadlines, with more urgent ones running at higher priorities. The Earliest Deadline First (EDF) algorithm is based on the dead line time constraint. The tasks were ordered in the increasing order of their deadlines and assigned to processors considering earliest deadline first.

In multiprocessor real time systems static algorithms are used to schedule periodic tasks whose characteristics are known a priori. Scheduling aperiodic tasks whose characteristics are not known a priori requires dynamic scheduling algorithms. Some researchers analyze the task scheduling problems based on the dynamic load balancing. It minimizes the execution time of single applications running in parallel on multi computer systems. It is essential for the efficient use of highly parallel systems with solving non uniform problems with unpredictable load estimates. In a distributed real time systems, uneven task arrivals temporarily overload some nodes and leave others idle or under loaded.

Power aware computing is not only for hand held devices that have limit energy supply, but also for large systems consisting of multiple processors (e.g., complex satellite and surveillance systems, data warehouses or web server farms), where the cost of energy consumption and cooling is substantial.

In the GA technique, the tasks are arranged as per their precedence level before applying GA operators. The cross over operator is applied for the tasks having different height and mutation operator is applied to the task having the same height. The fitness function only attempts to minimize processing time.

List scheduling techniques assign a priority to each task to be scheduled then sort the list of tasks in decreasing priority. As processors become available, the highest priority task in the task list is assigned to be processed and removed from the list. If more than one task has the same priority, selection from among the candidate tasks is typically random.

## 3. Task scheduling problem

Consider a directed acyclic task graph $G = \{V,E\}$ of $n$ vertices. Each vertex $V = \{T1, T2,...., Tn\}$ in the graph represents a task. Aim is to map every task to a set $P =$ $\{P1,P2, . . . , Pp\}$ of $p$ processors. Each task $Ti$ has a weight $Wi$ associated with it, which is the amount of time the task takes to execute on any one of the $p$ homogeneous processors. Each directed edge $eij$ indicates a dependency between the two tasks $Ti$ and $Tj$ that it connects. If there is a path from vertex $Ti$ to vertex $Tj$ in the graph $G$, then $Ti$ is the predecessor of $Tj$ and $Tj$ is the successor of $Ti$. The successor task cannot be executed before all its predecessors have been executed and their results are available at the processor at which the successor is scheduled to execute. A task is "ready" to execute on a processor if all of its predecessors have completed execution and their results are available at the processor on which the task is scheduled to execute. If the next task to be executed on a processor is not yet ready, the processor remains idle until the task is ready.

The goal is to find a schedule which is a mapping of tasks to processors that minimizes the makespan. The makespan of a schedule can be defined as the time it takes from the instant the first task begins execution to the instant at which the last task completes execution.

## 4 Scheduling algorithms

Scheduling is a key concept in computer multitasking and multiprocessing operating system design, and in real-time operating system design. In modern operating systems, there are typically many more processes running than there are CPUs available to run them. Scheduling refers to the way processes are assigned to run on the available CPUs. CPU scheduling deals with the problem of deciding which of the processes in the ready queue is to be allocated the CPU. There are many different CPU scheduling algorithms. This paper compares the performance of genetic algorithms with list scheduling heuristic.

### 4.1 Model

A simple task graph with 8 tasks is shown in Fig 1. The problem of optimal scheduling a task graph onto a multiprocessor system with p processors is to assign the computational tasks to the processors so that the precedence relations are maintained and all of the tasks are completed in the shortest possible time. The time that the last task is completed is called the finishing time (FT) of the schedule. Fig 2 shows a schedule for two processors displayers as Gantt chart. Fig 2 illustrates a schedule displayed as Gantt chart for the example task graph TG using two processors. This schedule has a finishing time of 10 units of time. An important lower bound for the finishing time of any schedule is the critical path length. The critical path length, $t_{cp}$ of a task graph is defined as the minimum time required completing all of the tasks in the task graph.

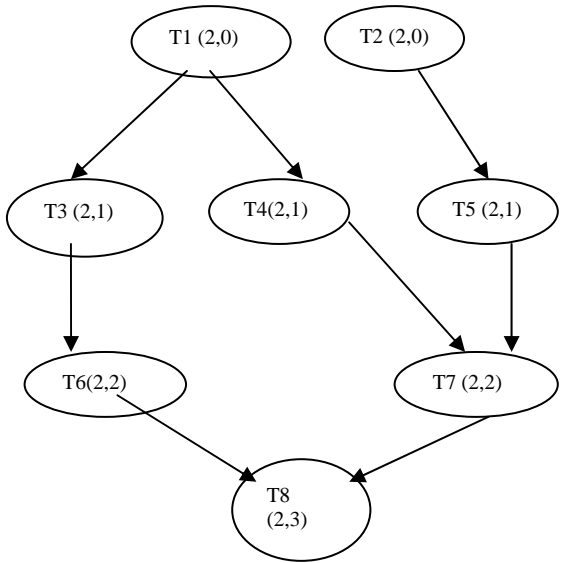

Fig 1 A task Graph TG

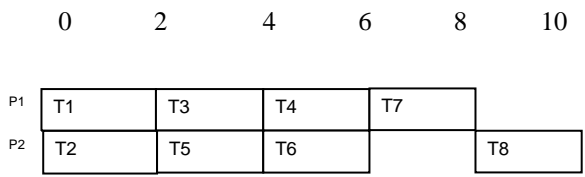

Fig 2 Gantt chart for scheduling two tasks

## 4.2 Genetic algorithms

Genetic algorithms try to mimic the natural evolution process and generally start with an initial population of individuals, which can either be generated randomly or based on some other algorithm. Each individual is an encoding of a set of parameters that uniquely identify a potential solution of the problem. In each generation, the population goes through the processes of crossover, mutation, fitness evaluation and selection. During crossover, parts of two individuals of the population are exchanged in order to create two entirely new individuals which replace the individuals from which they evolved. Each individual is selected for crossover with a probability of crossover rate. Mutation alters one or more genes in a chromosome with a probability of mutation rate. For example, if the individual is an encoding of a schedule, two tasks are picked randomly and their positions are interchanged. A fitness function calculates the fitness of each individual, i.e., it decides how good a particular solution is. In the selection process, each individual of the current population is selected into the new population with a probability proportional to its fitness. The selection process ensures that individuals with higher fitness values have a higher probability to be carried onto the next generation, and the individuals with lower fitness values are dropped out. The new population created in the above manner constitutes the next generation, and the whole process is terminated either after a fixed number of generations or when a stopping criteria is met. The population after a large number of generations is very likely to have individuals with very high fitness values which imply that the solution represented by the individual is good; it is very likely to achieve an acceptable solution to the problem. There are many variations of the general procedure described above. The initial population may be generated randomly, or through some other algorithm. The search space, i.e., the domain of the individuals, can be limited to the set of valid individuals, or extended to the set of all possible individuals, including invalid individuals. The population size, the number of generations, the probabilities of mutation and crossover are some of the other parameters that can be varied to obtain a different genetic algorithm.

The height varying point of the tasks are taken for cross over as shown in fig 3.After cross over the task arrangement is shown in fig 4. And their Gantt chart is shown in Fig 5.

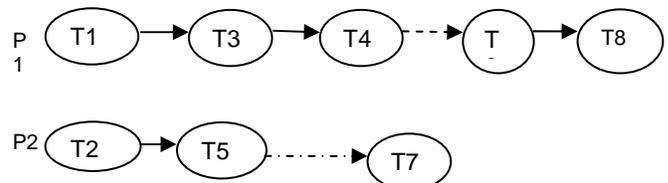

Fig 3 Cross over point

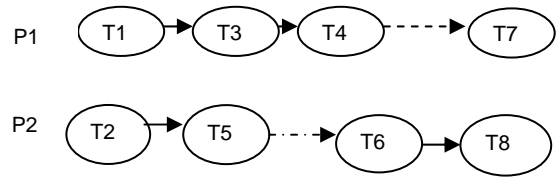

Fig 4 after Cross over

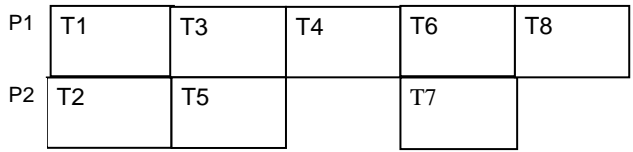

Fig 5 Gantt chart of tasks after Cross over.

## 4.3 List Scheduling Heuristic (LSH)

List scheduling techniques assign a priority to each task to be scheduled then sort the list of tasks in decreasing priority. As processors become available, the highest priority task in the task list is assigned to be processed and removed from the list. If more than one task has the same priority, selection from among the candidate tasks is typically random

## 5 Performance evaluations

### 5.1 Experimental setup

The genetic algorithm and List scheduling algorithm has been implemented and tested. Following are the assumptions under which the experiment is conducted. Assumptions about the task numbers are ranging from 8 to 110. The number of successors that each task node allowed is a random number between 3 and 6. The execution time for each task random number is assumed between 1 and 25. The task graphs are tested on a list-scheduling algorithm. The genetic algorithm used the following parameters throughout the simulation.
Population size = 20.
Maximum number of iterations = 500.
The simulation performed using MATLAB.

Table I Comparison of GA and LSH for various random task graph using two processors

| No.    of tasks | GA    finish time | LSH finish time |
|---|---|---|
| 8 | 11 | 10 |
| 17 | 22 | 24 |
| 28 | 43 | 47 |
| 33 | 48 | 61 |
| 39 | 46 | 60 |
| 44 | 50 | 56 |
| 49 | 55 | 103 |

From table 1 when the numbers of tasks are minimum of less than 15 LSH is the best solution to solve the scheduling problem

Table II Comparison of GA and LSH for various random task graph using three processors

| No. of tasks | GA finish time | LSH finish time |
|---|---|---|
| 23 | 26 | 28 |
| 28 | 43 | 47 |
| 33 | 48 | 61 |

Table III Comparison of GA and LSH for various random task graph using four processors

| No. of tasks | GA    finish time | LSH finish time |
|---|---|---|
| 39 | 46 | 60 |
| 44 | 50 | 56 |
| 49 | 55 | 103 |
| 54 | 68 | 68 |
| 59 | 83 | 121 |
| 69 | 104 | 139 |
| 79 | 137 | 153 |
| 89 | 141 | 157 |
| 99 | 158 | 211 |
| 100 | 158 | 222 |
| 110 | 174 | 222 |

From table 2 and 3 when the number of task and number of processors are increased GA gives the best solution. The time taken by the GA to compute the scheduling task is more than time taken by the List scheduling. But when the Number of Processors and number of tasks are increased the GA time is optimal with the List scheduling algorithm.

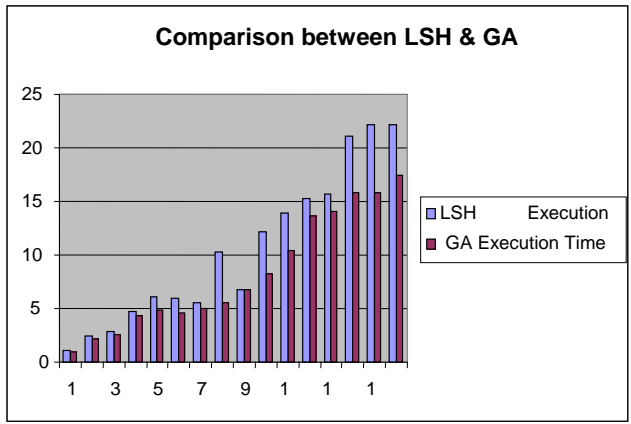

Fig 6 Comparison between LSH and GA

Table IV  Precedence relation Vs. Execution time

| Height | Best minimum time GA | LSH minimum time |
|--------|----------------------|------------------|
| 3 | 10 | 11 |
| 5 | 22 | 24 |
| 5 | 46 | 60 |
| 5 | 50 | 56 |
| 6 | 158 | 222 |
| 7 | 174 | 222 |

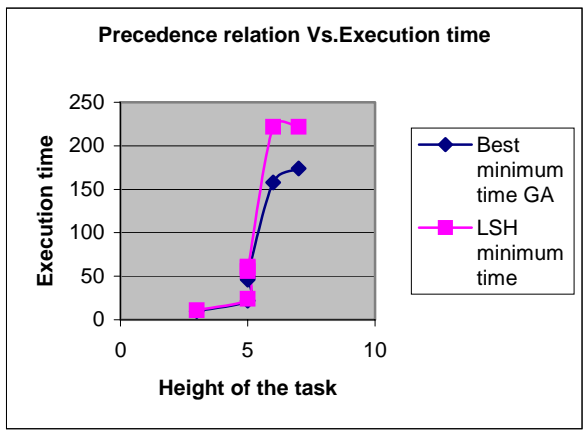

Fig 7: Precedence relation Vs. Execution time

From the fig 6 the LSH and GA produce almost same scheduling time when the number of tasks in range 8 to 28. When the numbers of tasks increase GA gives the better solution. When the tasks are more than 100, GA gives the best solution. One can infer from the table 4 and fig 7 the LSH and GA are producing the same result between the height 3 to 5.But when the height are more than 5 only GA produces the best result.

# 6. Conclusions

This paper compares scheduling algorithms for multiprocessor task scheduling .It conclude that from a purely performance point of view, Genetic algorithm is the best solution, but its deployment needs to be subject of a careful cost benefit analysis. It concludes that list scheduling will be suggested for the less number of tasks and processors.  It also concludes that the use of these algorithms are justified whenever the scheduling can be done off-line, there is a need for repeated execution of the schedules or the make span of the application is significantly longer than the scheduling time.

**Authors Profile**

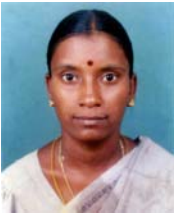

S.R.Vijayalakshmi is a Lecturer in School of Information Technology and Science, Dr.G.R.D college of science, Coimbatore. She received her B.Sc M.Sc, M.Phil in Electronics from the Bharathiar University and also received M.Sc in Computer Science from Bharathiar University and M.Phil in computer Science from Avinashilingam University for women. She has 14 years of experience in the teaching field. Her research interests include embedded systems, parallel and distributed systems, real time systems, real time operating systems and Microprocessors.

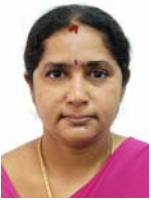

The author is a doctorate holder in Computer science with 21 years of experience in the academic side and approximately 2 years of experience in the industrial sector. She is the Professor and Head of the Department of Computer Science in Avinashilingam University for Women, Coimbatore-43. She has 80 publications at national and International level and executing funded projects worth 2 crores from UGC,AICTE and DRDO-NRB, DRDO-ARMREB. She is a life member of many professional organizations like CSI, ISTE, ISCA, WSEAS,AACE. Her areas of interest include network security, real time communication and real time operating systems. Her biography has been profiled at World's *Who's Who in Science and Engineering* Book, International Biographical Centre- Cambridge, England's - *Outstanding Scientist Worldwide for 2007, International Educator of the Year 2007* by IBC, *21st Century award for Achievement* by International Biographical Centre-Cambridge, England, *The International president's award for Iconic achievement*, by International Biographical Centre- Cambridge, England, *The Da Vinci Diamond for Inspirational Accomplishment* by International Biographical Centre, Cambridge .